\documentclass[aps,prd,superscriptaddress,twoside,twocolumn,nofootinbib,10pt,%
showpacs,floatfix]{revtex4-1}

\usepackage{amsmath,amssymb}
\usepackage{graphicx,bm}
\usepackage{slashed}
\usepackage{dcolumn}
\usepackage{epstopdf}
\usepackage{ulem} 
\usepackage[usenames]{color}

\allowdisplaybreaks

\begin{document}

\title{Dense baryonic matter in the hidden local symmetry approach:\\
Half-skyrmions and nucleon mass}

\author{Yong-Liang Ma}
\email{ylma@hken.phys.nagoya-u.ac.jp}
\affiliation{Department of Physics,  Nagoya University, Nagoya, 464-8602, Japan}

\author{Masayasu Harada}
\email{harada@hken.phys.nagoya-u.ac.jp}
\affiliation{Department of Physics,  Nagoya University, Nagoya, 464-8602, Japan}

\author{Hyun Kyu Lee}
\email{hyunkyu@hanyang.ac.kr}
\affiliation{Department of Physics, Hanyang University, Seoul 133-791, Korea}

\author{Yongseok Oh}
\email{yohphy@knu.ac.kr}
\affiliation{Department of Physics, Kyungpook National University, Daegu 702-701, Korea}

\author{Byung-Yoon Park}
\email{bypark@cnu.ac.kr}
\affiliation{Department of Physics, Chungnam National University, Daejeon 305-764, Korea}

\author{Mannque Rho}
\email{mannque.rho@cea.fr}
\affiliation{Department of Physics, Hanyang University, Seoul 133-791, Korea}
\affiliation{Institut de Physique Th\'eorique, CEA Saclay, 91191 Gif-sur-Yvette c\'edex, France}

\date{\today}
\begin{abstract}

Hadron properties in dense medium are treated in a unified way in a skyrmion model
constructed with an effective Lagrangian, in which the $\rho$ and $\omega$ vector mesons
are introduced as hidden gauge bosons, valid up to $O(p^4)$ terms in chiral expansion
including the homogeneous Wess-Zumino terms.
All the low energy constants of the Lagrangian -- apart from the pion decay constant and
the vector meson mass -- are fixed by the master formula derived from the relation between
the five-dimensional holographic QCD and the four-dimensional hidden local symmetry Lagrangian.
This Lagrangian allows one to pin down the density $n_{1/2}^{}$ at which the skyrmions
in medium fractionize into half-skyrmions, bringing in a drastic change in the equation of
state of dense baryonic matter.
We find that the $U(1)$ field that figures in the Chern-Simons term in the five-dimensional holographic
QCD action or equivalently the $\omega$ field in the homogeneous Wess-Zumino term in
the dimensionally reduced hidden local symmetry action plays a crucial role in the
half-skyrmion phase.
The importance of the $\omega$ degree of freedom may be connected to what happens in
the instanton structure of elementary baryon noticed in holographic QCD.
The most striking and intriguing in what is found in the model is that the pion decay constant
that smoothly drops with increasing density in the skyrmion phase stops decreasing at
$n_{1/2}^{}$ and remains nearly constant in the half-skyrmion phase.
In accordance with the large $N_c$ consideration, the baryon mass also stays nonscaling in
the half-skyrmion phase.
This feature which is reminiscent of the parity-doublet baryon model with a chirally invariant
mass $m_0^{}$ is supported by the nuclear effective field theory with the parameters of the
Lagrangian scaling modified at the skyrmion--half-skyrmion phase transition.
It also matches with one-loop renormalization group analysis based on hidden local symmetry.
A link between a nonvanishing $m_0^{}$ and the origin of nucleon mass distinctive from
dynamically generated mass is suggested.
We briefly discuss the possible consequences of the topology change found in this paper on
the forthcoming experiments at the rare isotope beam machines under construction.

\end{abstract}


\maketitle

\section{Introduction}

Understanding the QCD phase structure at high density is a wide-open problem in nuclear and
particle physics as it involves the highly nonperturbative region of strong interactions so far
unaccessible by lattice gauge calculations. Any theoretical investigation of the QCD phase structure, whether it is at the level of models
or effective field theories or more fundamental nature, would therefore be highly desirable.
In this paper, we investigate what happens to hadrons at high density in the theoretical framework
developed in Refs.~\cite{MOYHLPR12,MYOH12}.
To access dense baryonic matter, we exploit the approach proposed in Ref.~\cite{LPMRV03}, where
both the elementary baryons and multibaryon systems are described in a unified way with a single
effective mesonic Lagrangian with appropriate symmetries of QCD, the most prominent of which is
chiral symmetry, from which baryons are generated as solitons (called skyrmions) and many-body
interactions are taken into account with the given chiral Lagrangian.

In the past, for example in Ref.~\cite{PV09}, dense matter was studied with the Skyrme model
put on crystal lattice with pion fields only~\cite{Skyrme62} or minimally implemented with the
lowest-lying vector mesons $\rho$ and $\omega$ that will be referred to in what follows as
HLS$_{\rm min} (\pi,\rho,\omega)$%
\footnote{This terminology will be defined precisely below.}.
There are, however, no reasons to think that such models would succeed to describe adequately,
not to mention quantitatively, dense baryonic matter.
The Skyrme model has the leading current algebra term which can account for very low-energy
dynamics but the Skyrme quartic term that is crucial in stabilizing the soliton is neither unique
nor sufficient.
Given other degrees of freedom that can enter, there can be a variety of different
ways of stabilizing the soliton.
Even at the leading order in $N_c$, there is no reason why there should not be other quartic
as well as higher derivative terms.
But accounting for them -- which in principle should be feasible -- will bring in a plethora of
uncontrolled parameters, rendering the approach totally devoid of predictive power.
As for the minimal vector meson-implemented Skyrme model HLS$_{\rm min} (\pi,\rho,\omega)$,
the approximation used to simplify a part of the hidden local symmetry Lagrangian, namely
the homogeneous Wess-Zumino (hWZ) term, is unjustified when the vector meson
mass could decrease in density, which is the basic premise of hidden local symmetry (HLS)
theory~\cite{BKUY85,BKY88,HY03a}.

In this paper, we approach the problem with the five-dimensional Yang-Mills action in curved space plus the
Chern-Simons term that has been arrived at both dimensionally deconstructed, bottom-up,
starting from the current algebra term~\cite{SS03} and, top-down, from gravity-gauge duality
in string theory~\cite{SS04a,SS05}.%
\footnote{Note that the curvature is fixed by the gravity in the latter while undetermined in the
former. As will be specified, we will deal with a ``master formula" for any curvature.}
Kaluza-Klein-reduced to four dimensions, this theory consists of an infinite tower of hidden
local symmetric gauge fields.
Our assumption is that this is a theory of QCD type valid in the large $N_c$ -- and in the
case of the top-down model, large 't Hooft  $\lambda \equiv g^2 N_c$ -- limit.
One may ultimately use the five-dimensional action for dense baryonic matter as was done in
Refs.~\cite{HRYY07,HRYY07a,HSS08} for the single baryon generated as an instanton.
It is not yet known how to do this for dense baryonic matter although there has been an attempt
toward that direction~\cite{RSZ09,KMS12}.
As the first step, we follow the strategy of Refs.~\cite{HMY06,HMY10} to integrate out all but
the lowest vector mesons $\rho$ and $\omega$ of the infinite tower of vector and axial
vector mesons, making sure that hidden local symmetry is preserved up to $O (p^4)$ in
derivative expansion.
The structure of single baryon has been worked out in Refs.~\cite{MOYHLPR12,MYOH12} using
the holographic QCD model of Sakai and Sugimoto (SS for short)~\cite{SS04a,SS05} and also
with flat space in the SS model.
This brings about a drastic simplification from the infinite-tower theory but it has a great
advantage: Once the pion decay constant $f_\pi$ and the vector meson mass $m_\rho$ are
fixed from experiments, there are no undetermined parameters.
As explained in Ref.~\cite{MYOH12}, this comes about by the ``master formula" which allows us
to connect all the low-energy constants (LECs) of the (chiral) Lagrangian to the two constants
$f_\pi$ and $m_\rho$.

The most important feature that is found in Refs.~\cite{MOYHLPR12,MYOH12} was that the
$O (p^4)$ terms in the Lagrangian, in particular the hWZ terms that carry information on the
$\omega$ meson, could not be approximated by only a few terms -- not to mention ignored --
for the nucleon structure.
This observation suggested that the role of vector mesons, particularly, the $\omega$, would
be crucial in dense baryonic matter.
Indeed this will be what we will find in the present work.%
\footnote{As will be shown explicitly, the minimal version of the HLS model
HLS$_{\rm min} (\pi,\rho,\omega)$ exemplifies that it is a poor approximation to pick only
one term out of three that are in the hWZ action.}

The objective of this paper is to put the skyrmions constructed as described in
Refs.~\cite{MOYHLPR12,MYOH12} on the crystal lattice as was done for the Skyrme model.
Before stating our principal result, we should mention a possible shortcoming in our model.
When an HLS Lagrangian is implemented with baryons (that we shall refer to as BHLS),
the theory in principle should give at the mean field approximation a relativistic mean-field
model that describes nuclear matter properties as in the Walecka mean-field model.
However, as is well known, it is essential that there be a scalar meson in the theory giving rise
to the necessary attraction for binding~\cite{Song00}.
(See Ref.~\cite{PLRS13} and references given therein.)
Such a scalar, identified in Ref.~\cite{PLRS13} as ``dilaton," is missing in the HLS framework
used here.
How to introduce a scalar in the theory is highly problematic so we are ignoring it in this paper
but it may not be justified as we know from the original Skyrme model that a correct nucleon
mass can be obtained only when the Casimir contribution -- which is $O(N_c^0)$ and
captured in chiral perturbation theory at higher loop orders in the scalar channel -- is taken
into account.
As we will discuss, some of the attraction is generated by the $\omega$-$\rho$ coupling in
the crystal.
However the missing dilaton degree of freedom may still be needed for more
reliable calculations.
We will return to this matter in the last section.

The principal finding in this work is that the effective pion decay constant in medium
$f_\pi^*$ decreases smoothly up to the density $n_{1/2}^{}$ at which a skyrmion that has a baryon number one configuration fractionizes into two half-skyrmions each of which has half a baryon number, and then stays nonzero, more or less constant up to the chiral restoration density $n_c^{}$.
The notable consequence is that the effective nucleon mass $m_N^*$ also stops scaling for
$n>n_{1/2}^{}$. This is consistent with the large $N_c$ consideration.
In the large $N_c$ limit, a nucleon in medium as a skyrmion scales as $f_\pi^*$ and hence
remains more or less constant up to $n_c^{}$. What this means is that the nucleon mass could have a component that remains nonvanishing
up to the chiral transition, which is reminiscent of the parity doublet picture of the
baryon~\cite{DK89} with the nucleon mass given by
\begin{equation}
m_N^{} = m_0^{} +\Delta(\langle \bar{q}q \rangle),
\label{parity-doublet}
\end{equation}
where $m_0^{}$ is the chiral invariant mass and $\Delta$ is the part of the mass that vanishes
as $\langle \bar{q}q \rangle \rightarrow 0$ (in the chiral limit) for $n \rightarrow n_c^{}$.
The same result was obtained in a renormalization group (RG) analysis of hidden local symmetry
Lagrangian with baryons in Ref.~\cite{PLRS13} and also in a phenomenological study of the
equation of state (EoS) for dense compact-star matter in Ref.~\cite{DKLMR12}.

That $m_0^{} \neq 0$ implies that there is a part of the nucleon mass that is not
generated by spontaneous breaking of chiral symmetry.
The question then would be whether $m_0^{}$ reflects a fundamental feature of QCD as
suggested in Refs.~\cite{Gloz12,GLS12} or an emergent symmetry via correlations in medium
as in condensed matter as indicated in this crystal calculation.
This raises the issue of the origin of the nucleon mass as we shall discuss below.

This paper is organized as follows.
In Sec.~\ref{sec:sumhls} we briefly sketch the derivation of the HLS Lagrangian from
holographic models that we are using in the exploration of dense matter.
In Sec.~\ref{sec:fccmatter} we first discuss the main considerations in the investigation of the
dense skyrmion matter using the face-centered cubic (FCC) crystal in the framework of HLS,
then provide our numerical results of dense skyrmion matter and discuss their physical
implications.
We study the in-medium properties of hadrons with FCC crystal as the background in
Sec.~\ref{sec:hadron}.
In Sec.~\ref{matching}, we match our results to those obtained in (1) nuclear effective field
theory (EFT) focused on the EoS for compact star matter and (2) RG-analysis of HLS theory
with baryons.
Further remarks relevant to the physics of rare isotope beam (RIB) facilities and conclusions
are given in Sec.~\ref{sec:conclusion}.

\section{Hidden local symmetry and single skyrmion solution}
\label{sec:sumhls}

We briefly describe the HLS Lagrangian that we shall use in exploring dense matter.
The full symmetry group of the effective Lagrangian is
$G_{\rm full} = [ \mbox{SU(2)}_L \times \mbox{SU(2)}_R ]_{\rm chiral} \times
[\mbox{U(2)}]_{\rm HLS}$.
Here, $[\mbox{U(2)}]_{\rm HLS}$ is adopted as the ``hidden local symmetry" to incorporate
the $\rho$ and $\omega$ vector mesons.%
\footnote{Here for simplicity we are assuming that flavor $U(2)$ symmetry applies to
the $\rho$ and $\omega$ in medium as it seems to do in free space.
It is found by the RG analysis in Ref.~\cite{PLRS13} that the $U(2)$ must be broken down
appreciably in medium, so there may be a caveat in what we obtain here.}
The building blocks of the HLS Lagrangian are the two 1-forms $\hat{\alpha}_{\parallel \mu}$
and $\hat{\alpha}_{\perp \mu}$ defined by
\begin{eqnarray}
\hat{\alpha}_{\parallel \mu} &=&
\frac{1}{2i} (D_\mu \xi_R^{} \xi_R^\dagger + D_\mu \xi_L^{} \xi_L^\dagger), \nonumber\\
\hat{\alpha}_{\perp \mu} &=&
\frac{1}{2i} (D_\mu \xi_R^{} \xi_R^\dagger - D_\mu \xi_L^{} \xi_L^\dagger),
\end{eqnarray}
with the chiral fields $\xi_L^{}$ and $\xi_R^{}$, which are expressed in the unitary gauge as
\begin{eqnarray}
\xi_L^\dagger &=& \xi_R^{} = e^{i \pi/2f_\pi} \equiv \xi
\qquad \mbox{with} \qquad
\pi = \bm{\pi} \cdot \bm{\tau},
\end{eqnarray}
where $\bm{\tau}$'s are the Pauli matrices.
The covariant derivative is defined as
\begin{eqnarray}
D_\mu \xi_{R,L}^{} &=& (\partial_\mu - i V_\mu) \xi_{R,L}^{}
\end{eqnarray}
with $V_\mu$ being the gauge boson of the HLS.
This is the way to introduce vector mesons in the HLS, where the vector meson field
$V_\mu$ is~\cite{BKUY85,BKY88,HY03a}
\begin{eqnarray}
V_\mu = \frac{g}{2} \left( \omega_\mu + \rho_\mu \right)
\end{eqnarray}
with
\begin{eqnarray}
\rho_\mu = \bm{\rho}_\mu \cdot \bm{\tau} =
\left( \begin{array}{cc}
  \rho_\mu^0  & \sqrt{2} \rho_\mu^+ \\
  \sqrt{2} \rho_\mu^- &  -\rho_\mu^0
\end{array} \right) .
\end{eqnarray}

Up to $O(p^4)$ including the hWZ terms, the most general HLS Lagrangian can be expressed as
\begin{eqnarray}
\mathcal{L}_{\rm HLS} &=&
\mathcal{L}_{(2)} + \mathcal{L}_{(4)} + \mathcal{L}_{\rm anom} ,
\label{eq:Lag_HLS}
\end{eqnarray}
with
\begin{eqnarray}
\mathcal{L}_{(2)} &=&
f_\pi^2 \,\mbox{Tr}\, \left( \hat{\alpha}_{\perp\mu} \hat{\alpha}_{\perp}^{\mu} \right)
+ a f_\pi^2 \,\mbox{Tr}\, \left(\hat{\alpha}_{\parallel\mu} \hat{\alpha}_{\parallel}^{\mu} \right)
\nonumber\\ && \mbox{}
- \frac{1}{2g^2} \mbox{Tr}\, \left( V_{\mu\nu} V^{\mu\nu} \right),
\end{eqnarray}
where $f_\pi$ is the pion decay constant, $a$ is the parameter of the HLS,
$g$ is the vector meson coupling constant, and the field-strength tensor of
the vector meson is
\begin{eqnarray}
V_{\mu\nu} &=& \partial_\mu V_\nu - \partial_\nu V_\mu - i [V_\mu,V_\nu].
\end{eqnarray}
In the most general form of the $O(p^4)$ Lagrangian there are several terms
that include two traces in the flavor space such as the
$y_{10}^{}$--$y_{18}^{}$ terms listed in Ref.~\cite{HY03a}.%
\footnote{Another example of this kind is $\mbox{Tr}
\bigl[ \hat{\alpha}_\parallel^\mu \bigr] \mbox{Tr}
\bigl[ \hat{\alpha}_{\parallel\mu} \bigr]$ that generates the mass
difference between the $\rho$ and $\omega$ mesons.}
These terms are suppressed by $N_c$ compared to the other terms in the
Lagrangian and are not considered in the present work.
Then the $O(p^4)$ Lagrangian which we study in this paper is given by
\begin{eqnarray}
\mathcal{L}_{(4)} = \mathcal{L}_{(4)y} + \mathcal{L}_{(4)z} ,
\end{eqnarray}
where
\begin{widetext}
\begin{eqnarray}
\mathcal{L}_{(4)y} &=&
y_1^{} \mbox{Tr} \Bigl[ \hat{\alpha}_{\perp\mu}^{} \hat{\alpha}_\perp^\mu
\hat{\alpha}_{\perp\nu}^{} \hat{\alpha}_\perp^\nu \Bigr]
+ y_2^{} \mbox{Tr} \Bigl[ \hat{\alpha}_{\perp\mu}^{} \hat{\alpha}_{\perp\nu}^{}
\hat{\alpha}^\mu_\perp \hat{\alpha}^\nu_\perp \Bigr]
+ y_3^{} \mbox{Tr}
\left[ \hat{\alpha}_{\parallel\mu}^{} \hat{\alpha}_\parallel^\mu
\hat{\alpha}_{\parallel\nu}^{} \hat{\alpha}_\parallel^\nu \right]
+ y_4^{} \mbox{Tr}
\left[ \hat{\alpha}_{\parallel\mu}^{} \hat{\alpha}_{\parallel\nu}^{}
\hat{\alpha}^\mu_\parallel \hat{\alpha}^\nu_\parallel \right]
\nonumber \\ && \mbox{}
+ y_5^{} \mbox{Tr}
\left[ \hat{\alpha}_{\perp\mu}^{} \hat{\alpha}_\perp^\mu
\hat{\alpha}_{\parallel\nu}^{} \hat{\alpha}_\parallel^\nu \right]
+ y_6^{} \mbox{Tr}
\left[ \hat{\alpha}_{\perp\mu}^{} \hat{\alpha}_{\perp\nu}^{}
\hat{\alpha}^\mu_\parallel \hat{\alpha}^\nu_\parallel \right]
+ y_7^{} \mbox{Tr}
\left[ \hat{\alpha}_{\perp\mu}^{} \hat{\alpha}_{\perp\nu}^{}
\hat{\alpha}^\nu_\parallel \hat{\alpha}^\mu_\parallel \right]
\nonumber \\ && \mbox{}
+ y_8^{} \left\{
\mbox{Tr} \left[ \hat{\alpha}_{\perp\mu}^{} \hat{\alpha}_\parallel^\mu
\hat{\alpha}_{\perp\nu}^{} \hat{\alpha}_\parallel^\nu \right]
+ \mbox{Tr} \left[ \hat{\alpha}_{\perp\mu}^{} \hat{\alpha}_{\parallel\nu}^{}
\hat{\alpha}_\perp^\nu \hat{\alpha}_\parallel^\mu \right] \right\}
+ y_9^{} \mbox{Tr}
\left[ \hat{\alpha}_{\perp\mu}^{} \hat{\alpha}_{\parallel\nu}^{}
\hat{\alpha}^\mu_\perp \hat{\alpha}^\nu_\parallel \right],
\\
\mathcal{L}_{(4)z} & = &
i z_4^{} \mbox{Tr}
\Bigl[ V_{\mu\nu} \hat{\alpha}_\perp^\mu \hat{\alpha}_\perp^\nu \Bigr]
+ i z_5^{} \mbox{Tr}
\left[ V_{\mu\nu} \hat{\alpha}_\parallel^\mu \hat{\alpha}_\parallel^\nu \right].
\end{eqnarray}
\end{widetext}
Finally, the anomalous parity hWZ terms $\mathcal{L}_{\rm anom}$ are written as
\begin{eqnarray}
\Gamma_{\rm hWZ} & = & \int d^4x \mathcal{L}_{\rm anom} = \frac{N_c}{16\pi^2}\int_{M^4}
\sum_{i=1}^3 c_i \mathcal{L}_i ,
\end{eqnarray}
where $M^4$ stands for the four-dimensional Minkowski space and
\begin{subequations}
\begin{eqnarray}
\mathcal{L}_1 & = & i \, \mbox{Tr}\,
\bigl[ \hat{\alpha}_{\rm L}^3 \hat{\alpha}_{\rm R}^{}
 - \hat{\alpha}_{\rm R}^3 \hat{\alpha}_{\rm L}^{} \bigr], \\
\mathcal{L}_2 & = & i \, \mbox{Tr}\,
\bigl[ \hat{\alpha}_{\rm L}^{} \hat{\alpha}_{\rm R}^{}
\hat{\alpha}_{\rm L}^{} \hat{\alpha}_{\rm R}^{} \bigr]  ,  \\
\mathcal{L}_3 & = & \mbox{Tr}\,
\bigl[ F_{\rm V} \left( \hat{\alpha}_{\rm L}^{} \hat{\alpha}_{\rm R}^{}
 - \hat{\alpha}_{\rm R}^{} \hat{\alpha}_{\rm L}^{} \right) \bigr] ,
\end{eqnarray}
\end{subequations}
in the 1-form and 2-form notations with
\begin{eqnarray}
\hat{\alpha}_{L}^{} & = & \hat{\alpha}_\parallel^{} - \hat{\alpha}_\perp^{},
\nonumber \\
\hat{\alpha}_{R}^{} & = & \hat{\alpha}_\parallel^{} + \hat{\alpha}_\perp^{},
\nonumber \\
F_V & = & dV - i V^2.
\end{eqnarray}

As for the low energy constants $f_\pi$, $a$, $g$, $y_i$, $z_i$, and $c_i$, we use the
general ``master formula" proposed in Refs.~\cite{MOYHLPR12,MYOH12} from a class of
holographic QCD models, namely,
\begin{eqnarray}
f_{\pi}^2 & = & N_c G_{\rm YM}^{} M_{KK}^2  \int dz K_2(z)
\left[ \dot{\psi}_0^{}(z)
\right]^2, \nonumber\\
a f_{\pi}^2 & = & N_c G_{\rm YM}^{} M_{KK}^2
\lambda_1^{} \langle \psi^2_1 \rangle,
\nonumber\\
\frac{1}{g^2} & = & N_c G_{\rm YM}^{} \langle \psi_1^2 \rangle ,
\nonumber\\
y_1^{} & = & -y_2^{} = -N_c G_{\rm YM}^{} \left\langle
\left(1 + \psi_1 - \psi_0^2 \right)^2
\right\rangle ,
\nonumber\\
y_3^{} & = & -y_4^{} = -N_c G_{\rm YM}^{} \left\langle
\psi^2_1 \left(1 + \psi_1^{} \right)^2
\right\rangle ,
\nonumber\\
y_5^{} & = & 2 y_8^{} = -y_9^{} = -2N_c G_{\rm YM}^{}
\left\langle \psi_1^2 \psi_0^2 \right\rangle ,
\nonumber\\
y_6^{} & = & - \left( y_5^{} + y_7^{} \right) ,
\nonumber\\
y_7^{} & = & 2N_c G_{\rm YM}^{} \left\langle \psi_1^{}
\left ( 1 + \psi_1^{} \right)
\left(1 + \psi_1^{} - \psi_0^2 \right) \right\rangle ,
\nonumber\\
z_4^{} & = & 2N_c G_{\rm YM}^{} \left\langle \psi_1^{}
\left( 1+\psi_1^{} - \psi_0^2 \right)
\right\rangle ,
\nonumber\\
z_5^{} & = & -2N_c G_{\rm YM}^{} \left\langle \psi_1^2
\left( 1 + \psi_1^{} \right) \right\rangle ,
\nonumber\\
c_1^{} & = &  \left\langle\!\!\left\langle
\dot{\psi}_0^{} \psi_1^{} \left( \frac{1}{2} \psi_0^2 + \frac{1}{6} \psi_1^2
- \frac{1}{2} \right) \right\rangle\!\!\right\rangle ,
\nonumber\\
c_2^{} & = & \left\langle\!\!\left\langle
\dot{\psi}_0^{} \psi_1^{} \left( -\frac{1}{2} \psi_0^2 + \frac{1}{6} \psi_1^2
+ \frac{1}{2} \psi_1^{} + \frac{1}{2} \right) \right\rangle\!\!\right\rangle,
\nonumber\\
c_3^{} & = & \left\langle\!\!\left\langle
\frac{1}{2}\dot{\psi}_0^{} \psi_1^{2} \right\rangle\!\!\right\rangle ,
\label{eq:lecshls}
\end{eqnarray}
where the smallest nonzero eigenvalue $\lambda_1^{}$ and its corresponding eigenfunction
$\psi_1^{}(z)$ satisfy the eigenvalue equation
\begin{eqnarray}
- K_1^{-1}(z)\partial_z \left[ K_2(z) \partial_z \psi_n^{} (z) \right]
&=& \lambda_n^{} \psi_n^{}(z),
\end{eqnarray}
with $K_1(z)$ and $K_2(z)$ being the warping factors of the five-dimensional space-time.
In Eq.~(\ref{eq:lecshls}) we have made the following definitions:
\begin{eqnarray}
\langle A \rangle & \equiv & \int_{-\infty}^{\infty}  dz K_1(z) A(z),
\nonumber\\
\langle\langle A \rangle\rangle &\equiv & \int_{-\infty}^\infty dz A(z).
\end{eqnarray}

In the present work, we take as in Refs.~\cite{MOYHLPR12,MYOH12} $K_1(z) = K^{-1/3}(z)$
and $K_2(z) = K(z)$ with $K(z) = 1 + z^2$ corresponding to the Sakai-Sugimoto
model~\cite{SS04a,SS05}.
Therefore, once the hQCD parameters $M_{\rm KK}$ and $G_{\rm YM}$ are determined by
using the real world parameters, for example, such as $f_\pi$ and $m_\rho$, all the rest of the parameters
of the effective Lagrangian except the parameter $a$ are fixed through the master formula given by Eq.~(\ref{eq:lecshls}).
In this work, we will work with the parameter set determined by the empirical values,
\begin{eqnarray}
f_\pi = 92.4 \mbox{ MeV}, \qquad m_\rho = 775.5 \mbox{ MeV}.
\label{PMS}
\end{eqnarray}

Note that due to the relations between $y_i^{}$'s, the $\omega$ meson can couple to the others only
through the hWZ terms.
We will denote the $\omega$-decoupled HLS Lagrangian with $c_{i}^{}=0$ for all $i$ as
$\mbox{HLS}(\pi, \rho)$ in order to distinguish it from the full Lagrangian
$\mbox{HLS}(\pi, \rho, \omega)$.
If we keep only the terms up to $O(p^2)$ and only one term among the hWZ action that
results if one takes the equation of motion for the $\rho$ meson and the large $m_\rho$
limit in the hWZ action in Eq.~(\ref{eq:Lag_HLS})%
\footnote{As mentioned, this limit is at odds with the basic premise of HLS~\cite{HY03a}.}
and use the same values for the parameters, the Lagrangian is reduced to the
``minimum model" adopted in Ref.~\cite{PRV03}, which is denoted as
$\mbox{HLS}_{\rm min}(\pi, \rho, \omega)$.

In Refs.~\cite{MOYHLPR12,MYOH12}, the single skyrmion solution was found under the hedgehog
configuration for various parameter sets.
For $\mbox{HLS}(\pi, \rho, \omega)$ with the parameters determined by Eq.~(\ref{PMS}),
the soliton mass $M_{\rm sol}$ is 1184~MeV and the root mean square winding number
radius $\sqrt{\langle r^2 \rangle_W}$ is 0.433~fm.
Compared to the $\mbox{HLS}_{\rm min}(\pi, \rho, \omega)$ which yields
$M_{\rm sol} = 1407$~MeV and $\sqrt{\langle r^2 \rangle_W}=0.540$~fm, we can see that
the $O(p^4)$ terms provide about 300~MeV reduction in the soliton mass and about
$0.1$~fm reduction in the size.
This means that the $\pi$-$\rho$ interactions incorporated through the $O(p^4)$ terms and
the $\pi$-$\rho$-$\omega$ interactions through the hWZ terms provide such an effective
attraction.
As we will see later, the reduction in the size of the skyrmion plays the most important role in
determining the critical density $n_{1/2}^{}$, where the half-skyrmion phase transition happens.

On the other hand, if the $\omega$ is decoupled, $M_{\rm sol}$ is reduced to 834~MeV and
the skyrmion has a very small size as $\sqrt{\langle r^2 \rangle_W} = 0.247$~fm.
This emphasizes the role of the repulsive $\omega$ meson in the HLS Lagrangian.

\section{Skyrmion matter from the hidden local symmetry}
\label{sec:fccmatter}

\begin{table*}[t]
\caption{Symmetries of the FCC skyrmion crystal.}\label{table:fccsymmetry}
\centering
\begin{tabular}{cccccc}
\hline\hline
&
& $\begin{array}{c}\mbox{Reflection}\\ \mbox{($yz$ plane)}\end{array}$
& $\begin{array}{c}\mbox{threefold axis}\\ \mbox{rotation}\end{array}$
& $\begin{array}{c}\mbox{fourfold axis}\\ \mbox{($z$ axis) rotation}\end{array}$
& Translation \\
\hline
$(x,y,z)$
& $\rightarrow$
& $(-x,y,z)$
& $(y,z,x)$
& $(x,z,-y)$
& $(x+L,y+L,z)$ \\
$U=\sigma+i\bm{\tau} \cdot \bm{\pi}$
& $\rightarrow$
& $(\sigma, -\pi_1, \pi_2, \pi_3)$
& $(\sigma, \pi_2, \pi_3, \pi_1)$
& $(\sigma, \pi_1, \pi_3, -\pi_2)$
& $(\sigma, -\pi_1, -\pi_2, \pi_3)$ \\
$\rho^a_i \equiv \varepsilon_{aip} \tilde{\rho}_p $
& $\rightarrow$
& $(-\tilde{\rho}_1,\tilde{\rho}_2,\tilde{\rho}_3)$
& $(\tilde{\rho}_2,\tilde{\rho}_3,\tilde{\rho}_1)$
& $(\tilde{\rho}_1,\tilde{\rho}_3,-\tilde{\rho}_2)$
& $(-\tilde{\rho}_1,-\tilde{\rho}_2,\tilde{\rho}_3)$ \\
$\omega_0$
& $\rightarrow $
& $\omega_0$
& $\omega_0$
& $\omega_0$
& $\omega_0$ \\
\hline\hline
\end{tabular}
\end{table*}

Since the skyrmions are classical solitons, they provide a natural framework to study dense matter.
We can make the dense skyrmion matter by putting more and more skyrmions into the space
without worrying about such a serious problem as the sign problem in the lattice QCD.
We shall consider an FCC crystal made up of skyrmions. Such crystal configuration is known to give the lowest ground-state energy among the crystal symmetries studied so far~\cite{KS88,KS89a}.
In the lowest energy configuration, the skyrmion at each lattice site should be arranged in such
a way that the nearest skyrmions have the maximum attraction.
In the study with the original Skyrme model, it turns out that two skyrmions apart from each other
have the maximum attraction when one of them is rotated in the isospin space with respect to the
other by an angle $\pi$ about the axis perpendicular to the line joining them.
In Ref.~\cite{PRV03}, it was shown in a simple approximation that such a configuration is also
most attractive in the model of $\mbox{HLS}_{\rm min}(\pi,\rho,\omega)$.
If we assume that it holds in the case of $\mbox{HLS}(\pi,\rho)$ and
$\mbox{HLS}(\pi,\rho,\omega)$, then the FCC crystal configuration for the $\pi$, $\rho$, and
$\omega$ should have the symmetries with respect to the reflection and translations as
summarized in Table~\ref{table:fccsymmetry}.

The classical solutions for $\pi$, $\rho$, and $\omega$ mesons satisfying the symmetries
for the maximum attraction and the constraints for each FCC box to carry a given baryon
number can be obtained by using the Fourier expansion method developed in
Refs.~\cite{KS88,KS89a} for the original Skyrme model with the pions only,
then generalized in Ref.~\cite{PRV03} for the model with vector mesons.
The main idea is to introduce the unnormalized fields $\bar{\phi}^{M}_\mu$
($M = \pi,\rho$), each of which can be expanded in terms of the Fourier series as
\begin{eqnarray}
\bar{\phi}^M_0 & = &
\displaystyle
\sum_{abc} \beta^M_{abc} \cos(\pi ax/L)\cos(\pi by/L)\cos(\pi cz/L),
\nonumber\\
\bar{\phi}^M_1 & = &
\displaystyle
\sum_{hkl} \alpha^M_{hkl} \sin(\pi hx/L)\cos(\pi ky/L)\cos(\pi lz/L),
\nonumber\\
\bar{\phi}^M_2 &=&
\displaystyle
\sum_{hkl} \alpha^M_{hkl} \cos(\pi lx/L)\sin(\pi hy/L)\cos(\pi kz/L),
\nonumber\\
\bar{\phi}^M_3 &=&
\displaystyle
\sum_{hkl} \alpha^M_{hkl} \cos(\pi kx/L)\cos(\pi ly/L)\sin(\pi hz/L),
\nonumber \\
\label{eq:piexpansion}
\end{eqnarray}
and thereafter normalized as
\begin{equation}
\phi^M_\alpha  =  \frac{\bar{\phi}^M_{\alpha}}
{\sqrt{\sum_{\beta}\left(\bar{\phi}^M_\beta\right)^2}},
\end{equation}
for each $M=\pi, \rho$.
Here, $2L$ is the size of a single FCC box containing four skyrmions inside.
Thus, the skyrmion number density $n$ is given by $n=1/(2L^3)$.
If we take skyrmions as nucleons, the normal nuclear matter density
$n_0^{} = 0.17/\mbox{fm}^3$ corresponds to $L \sim 1.43$~fm.

Finally, $U(\equiv \xi^2)$ and $\rho^a_i$ can be expressed in terms of these
normalized fields as
\begin{eqnarray}
U = \phi^\pi_0 + i \bm{\tau} \cdot \bm{\phi}^\pi, \qquad
g \rho^a_i =  \frac{\varepsilon_{abc} {\phi}^\rho_b \partial_i {\phi}^\rho_c}{1+\phi^\rho_0}.
\end{eqnarray}

The isoscalar field $\omega_0^{}$ has the same symmetry properties as the $\phi^\pi_0$ and
therefore could be expanded as
\begin{eqnarray}
\omega_0^{} = \sum_{abc} \beta^\omega_{abc} \cos(\pi ax/L)\cos(\pi by/L)\cos(\pi cz/L)
\end{eqnarray}
with the expansion coefficients $\beta_{abc}^\omega$.

The Fourier coefficients and their indices are subject to the topological structure and the
constraints from the symmetry properties listed in Table~\ref{table:fccsymmetry}.
From the symmetry properties one can easily check that in the above equations
$(a,b,c)$ should be all even or all odd integers, and $(k,l)$ should be all even (odd) if $h$ is odd (even).
Furthermore, to provide the correct topological structure (winding number) to the configuration,
the expansion coefficients $\beta^\pi_{abc}$ must satisfy the constraint,
\begin{eqnarray}
\sum_{\mbox{\scriptsize even}}\beta^\pi_{abc} = 0.
\end{eqnarray}

\begin{widetext}
As for the pion and $\rho$ field configurations, all the Fourier coefficients are determined
by minimizing the skyrmion energy per baryon, $E/B$, which, from the
Lagrangian (\ref{eq:Lag_HLS}), is expressed as
\begin{eqnarray}
E/B  =  (E/B)_{O(p^2)} + (E/B)_{O(p^4)} + (E/B)_{\rm hWZ},
\label{EperB}
\end{eqnarray}
where
\begin{eqnarray}
(E/B)_{O(p^{2})} &=& \frac{1}{4} \int_{\rm Box} d^3x \left[ \frac{f_\pi^2}{2}
\left( \tilde{\bm{\alpha}}_{\perp i} \cdot \tilde{\bm{\alpha}}_{\perp i} \right) - \frac{m_\omega^2}{2} \omega^2 +
\frac{m_\rho^2}{2g^2} \left( \tilde{\bm{\alpha}}_{\parallel i} \cdot \tilde{\bm{\alpha}}_{\parallel i} \right)
- \frac{1}{2} \partial_i^{} \omega_0^{} \partial_i^{} \omega_0^{}
+ \frac{1}{g^2} \bm{V}_{ij} \cdot \bm{V}_{ij} \right] ,
\\
(E/B)_{O(p^{4})} &=&
 - \frac{1}{4} \int_{\rm Box} d^3x \Biggl[
\frac{y_1^{}}{4} \left( \tilde{\bm\alpha}_{\perp i} \times \tilde{\bm\alpha}_{\perp j} \right)
\cdot \left( \tilde{\bm\alpha}_{\perp i} \times \tilde{\bm\alpha}_{\perp j} \right)
+ \frac{y_3^{}}{4} \left( \tilde{\bm\alpha}_{\parallel i} \times \tilde{\bm\alpha}_{\parallel j} \right)
\cdot \left( \tilde{\bm\alpha}_{\parallel i} \times \tilde{\bm\alpha}_{\parallel j} \right)
\nonumber\\
&&
\hskip 6em \mbox{}
+ \frac{y_5^{}}{4} \left\{
\left(\tilde{\bm\alpha}_{\perp i} \times \tilde{\bm\alpha}_{\parallel j} \right) \cdot
\left(\tilde{\bm\alpha}_{\perp i} \times \tilde{\bm\alpha}_{\parallel j} \right)
- \left(\tilde{\bm\alpha}_{\perp i} \times \tilde{\bm\alpha}_{\parallel i} \right) \cdot
\left(\tilde{\bm\alpha}_{\perp j} \times \tilde{\bm\alpha}_{\parallel j} \right) \right\}
\nonumber\\
&&
\hskip 6em \mbox{}
  + \frac{y_6^{}}{4} \left( \tilde{\bm\alpha}_{\perp i} \times \tilde{\bm\alpha}_{\perp j} \right)
\cdot \left( \tilde{\bm\alpha}_{\parallel i} \times \tilde{\bm\alpha}_{\parallel j} \right)
- \frac{z_4^{}}{2} \bm{V}_{ij} \cdot \left( \tilde{\bm\alpha}_{\perp i }
\times \tilde{\bm\alpha}_{\perp j}\right)
- \frac{z_5^{}}{2} \bm{V}_{ij} \cdot \left(
\tilde{\bm\alpha}_{\parallel i} \times \tilde{\bm\alpha}_{\parallel j}\right) \Biggr] ,
\\
(E/B)_{\rm hWZ}  &=& \frac{1}{4} \left( \frac{gN_c}{32\pi^2}\right)
\int_{\rm Box} d^3x \Biggl[ -c_1^{} \omega_0^{} \varepsilon_{ijk}
\left[ \tilde{\bm\alpha}_{\perp i} \cdot \left( \tilde{\bm\alpha}_{\parallel j} \times
\tilde{\bm\alpha}_{\parallel k} \right)
+ \tilde{\bm\alpha}_{\perp i} \cdot \left( \tilde{\bm\alpha}_{\perp j} \times
\tilde{\bm\alpha}_{\perp k } \right) \right]
\nonumber\\
&& \hskip 9em \mbox{}
 - c_2^{} \omega_0^{} \varepsilon_{ijk} \left[ \tilde{\bm\alpha}_{\perp i} \cdot
\left(\tilde{\bm\alpha}_{\parallel j} \times \tilde{\bm\alpha}_{\parallel k} \right)
- \tilde{\bm\alpha}_{\perp i} \cdot \left( \tilde{\bm\alpha}_{\perp j} \times
\tilde{\bm\alpha}_{\perp k}\right)\right]
\nonumber \\
&& \hskip 9em \mbox{}
+ 2 c_3^{} \varepsilon_{ijk}
\left[ \partial_i \omega_0 \left( \tilde{\bm\alpha}_{\parallel j} \cdot
\tilde{\bm\alpha}_{\perp k} \right)
+ \omega_0 \left( \bm{V}_{ij} \cdot \tilde{\bm\alpha}_{\perp k} \right) \right] \Biggr]
\end{eqnarray}
\end{widetext}
with the integration being taken over the space inside an FCC box of size $L$. Here, 
$\tilde{\bm\alpha}_{\perp i}$, $\tilde{\bm\alpha}_{\parallel i}$, and $\bm{V}_{ij}$ are
expressed in terms of $\phi^{\pi,\rho}_\mu$ as
\begin{eqnarray}
\tilde{\alpha}_{\perp i}^a
&=&
\left[\partial_i \phi^{\pi} - \frac{\phi^\pi}{1 + \phi^\pi_0} \partial_i \phi^\pi_0 \right]^a,
\\
\tilde{\alpha}_{\parallel i}^a
&=&
\left[ \frac{(\partial_i \phi^\pi \times \phi^\pi )}{(1+\phi^\pi_0)}
- \frac{(\partial_i \phi^\rho \times \phi^\rho )}{(1+\phi^\rho_0)} \right]^a,
\\
V_{ij}^a &=&
\frac12 \Biggl\{
- \frac{\partial_i \phi^\rho_0}{1+\phi^\rho_0} (g \rho^a_j)
+ \frac{\partial_j \phi^\rho_0}{1+\phi^\rho_0} (g \rho^a_i)
\nonumber\\
&& \quad\!\!\!\! \mbox{}
- 2 \frac{(\partial_i \phi^\rho \times \partial_j \phi^\rho )^a}{1 + \phi^\rho_0}
+ \left[ (g \rho_i) \times (g\rho_j) \right]^a
\Biggr\}.
\end{eqnarray}

On the other hand, the simple minimization procedure of $E/B$ to find the expansion
coefficients $\beta^{\omega}_{abc}$ would lead to the trivial solution $\omega_0^{} = 0$.
This is because there are no constraints for the $\beta^\omega_{abc}$'s, and the $\omega$
always provides repulsive interactions.
Therefore,  the $\omega$ expansion coefficients should be determined by satisfying the
equation of motion for $\omega (=\omega_0^{}/f_\pi)$; {\it viz.}
\begin{eqnarray}
\left( - \partial_i \partial_i + m_\omega^2\right)\omega(r) &=&
J^\omega - \partial_i J_i^\omega  ,
\end{eqnarray}
where $J^\omega$ and $\partial_i J_i^\omega$ coming from the hWZ terms are
\begin{eqnarray}
J^\omega &=&  - \frac{gN_c}{32\pi^2} \varepsilon_{ijk}
\Bigl[ (c_1 + c_2) \, \tilde{\bm{\alpha}}_{\perp i } \cdot
\left(\tilde{\bm{\alpha}}_{\parallel j} \times \tilde{\bm\alpha}_{\parallel k} \right) \nonumber\\
&&  \qquad \qquad \mbox{}
+ (c_1 - c_2)\tilde{\bm\alpha}_{\perp i }\cdot\left(
\tilde{\bm\alpha}_{\perp j } \times \tilde{\bm\alpha}_{\perp k }\right) \nonumber\\
&& \qquad \qquad \mbox{}
- 2 c_3 \bm{V}_{ij} \cdot \tilde{\bm{\alpha}}_{\perp k } \Bigr] , \nonumber\\
J^\omega_i & = &  \frac{gN_c}{16\pi^2} \, c_3^{}\, \varepsilon_{ijk} \,
\tilde{\bm\alpha}_{\parallel j} \cdot \tilde{\bm\alpha}_{\perp k}  .
\end{eqnarray}
Due to the crystal structure, the source term can be expanded in terms of the Fourier series as
\begin{equation}
J^\omega - \partial_i J_i^\omega =
\sum_{abc} \gamma_{abc}^{} \cos(\pi ax/L)\cos(\pi by/L)\cos(\pi cz/L).
\end{equation}
Then, thanks to the linear nature of the equation of motion for the $\omega$,
$\beta^\omega_{abc}$ can be easily obtained as
\begin{equation}
\beta^\omega_{abc} = \frac{\gamma_{abc}}{(\pi/L)^2(a^2+b^2+c^2) +m_{\omega}^2}.
\end{equation}
Thus, in the numerical minimizing processes only the $\alpha_{hkl}^{\pi, \rho}$ and
$\beta^{\pi,\rho}_{abc}$ are taken as adjustable parameters.

\begin{figure}[t]
\centering
\includegraphics[scale=0.5]{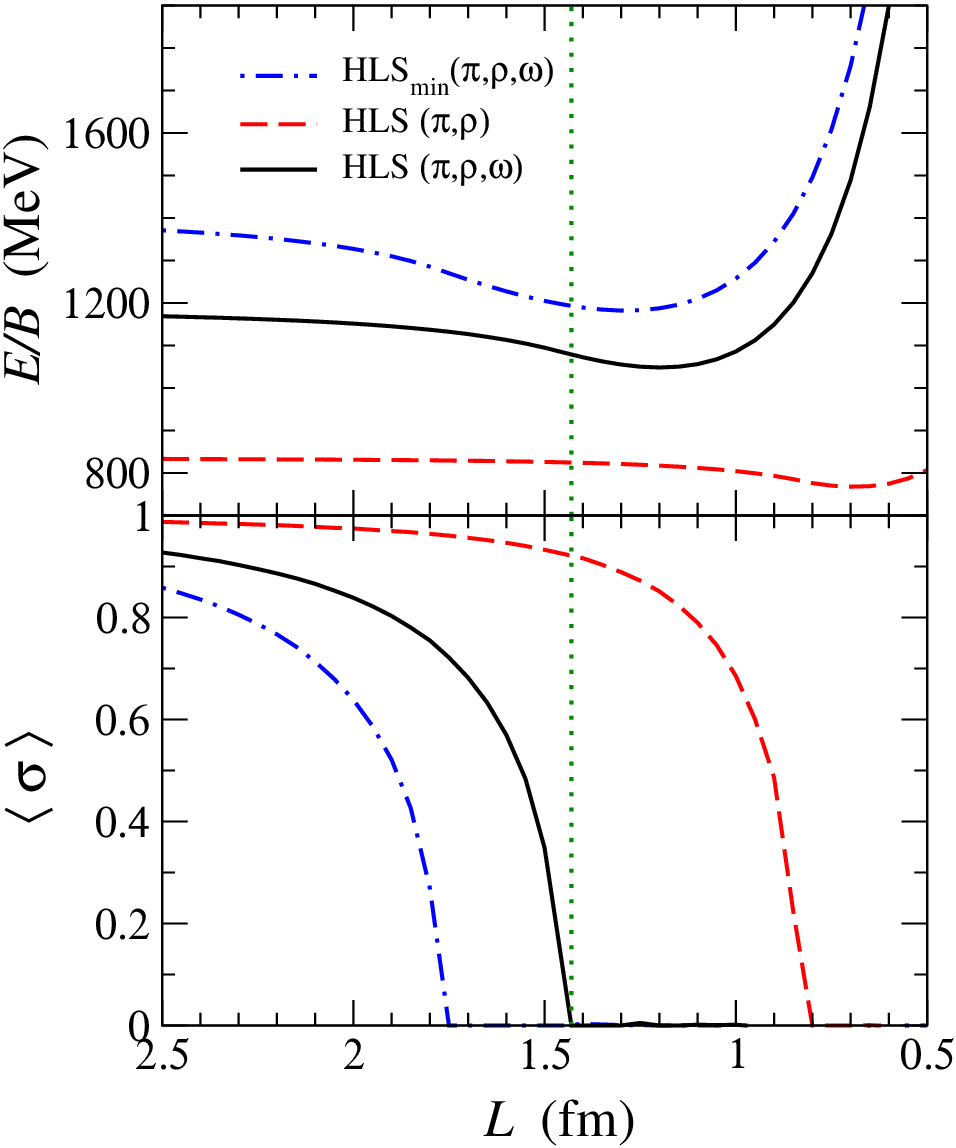}
\caption[]
{$E/B$ and $\langle \sigma \rangle$ of the minimum energy configuration for a given $L$.
Here and in all the figures that follow, $L$ decreases to the right, indicating increase
in density.
The numerical results for HLS$(\pi,\rho,\omega)$, HLS$(\pi,\rho)$, and
HLS$_{\rm min}(\pi,\rho,\omega)$ are presented by solid, dashed, and
dash-dotted lines, respectively. }
\label{fig:compareSSmin}
\end{figure}

Shown in Fig.~\ref{fig:compareSSmin} are the numerical results for the dependence of
$E/B$ and $\langle \sigma\rangle$ on the FCC size parameter $L$.
Here, $\langle \sigma \rangle$ is the average of the chiral field $U$ over the space
that a single skyrmion occupies and it vanishes in the half-skyrmion phase.
Here, we present the numerical results for HLS$(\pi,\rho,\omega)$, HLS$(\pi,\rho)$
and HLS$_{\rm min}(\pi,\rho,\omega)$ by solid, dashed, and dash-dotted lines, respectively.
The position of the normal nuclear matter density is denoted by a vertical dotted line.

One can see that in HLS$(\pi,\rho,\omega)$, the half-skyrmion phase transition
takes place close to the normal nuclear matter density $n_0^{}$.
Comparing to that of the minimal model HLS$_{\rm min}(\pi,\rho,\omega)$, we see
that the inclusion of the $O(p^4)$ interaction terms and the other $\pi$-$\rho$-$\omega$
interactions through the hWZ  terms makes $n_{1/2}^{}$ larger by a factor of $1.7$,
which is quite a noticeable effect.
This higher value for $n_{1/2}^{}$ in HLS$(\pi,\rho,\omega)$ compared to that of the
minimal model may come from the fact that the size of the single skyrmion is smaller
in the former and that the additional interactions in HLS$(\pi,\rho,\omega)$ weaken
the repulsive interactions from the $\omega$.
This analysis is supported by the results from HLS$(\pi, \rho)$, where the half-skyrmion
phase transition occurs at $L \sim 0.8$~fm, i.e., $n_{1/2}^{} \sim 6 \, n_0$, much higher
than with the $\omega$.
This also accounts that in HLS$(\pi, \rho)$, the absence of the $\omega$ reduces
the skyrmion size to nearly half of that of HLS$(\pi, \rho, \omega)$.

The density $n_{\rm min}^{}$ at which $E/B$ is the minimum turns out to be larger
than $n_{1/2}^{}$ for all three models considered in the present work.
As one can see from Fig.~\ref{fig:compareSSmin}, the minimum is achieved at
$n_{\rm min}^{} / n_{1/2}^{} \sim 2$ in HLS$(\pi,\rho,\omega)$ and in
HLS$_{\rm min}(\pi, \rho,\omega)$, but at $n_{\rm min}^{} / n_{1/2}^{} \sim 1.5$ in
HLS$(\pi,\rho)$.
The binding energy per baryon at this minimum energy configuration can be also read
from Fig.~\ref{fig:compareSSmin} as $\sim 150$~MeV, $\sim 100$~MeV, and
$\sim 50$~MeV for HLS$_{\rm min}(\pi, \rho,\omega)$, HLS$(\pi, \rho,\omega)$, and
HLS$(\pi, \rho)$, respectively.
As a complete theory of nuclear matter, the minimum at $n_{\rm min}^{}$ should represent
the nuclear matter ground state.
That $n_{\rm min}^{} >n_0^{}$ and the binding energy at $n_{\rm min}^{}$ is larger than
the empirical value $\sim 16$~MeV per baryon signals that there is something missing in
the present skyrmion crystal description for the EoS of nuclear matter.
We will discuss the plausible causes for this defect and possible remedies in the last section.

It is of theoretical, though perhaps academic, interest that the $E/B$ versus density of
HLS$(\pi, \rho)$ is (nearly) flat.
One can easily understand this feature, which arises from the important role of the
vector mesons in nucleon structure.
In Refs.~\cite{Sutcliffe10,Sutcliffe11}, the five-dimensional BPS theory -- which can be
considered as the Sakai-Sugimoto hQCD model with flat metric -- is transformed to
a four-dimensional infinite-tower meson theory and the instanton in that theory that
satisfies the Bogomolnyi bound condition is then transformed to the skyrmion.
It is shown that, when only the pions are considered, the skyrmion carries an excess energy
of about $25$\% relative to the Bogomolnyi bound.
However, when the lowest $SU(2)$ vector meson, the $\rho$ meson, is included, the energy
is drastically reduced, arriving almost at the Bogomolnyi bound.
The next vector meson, which is the axial $a_1^{}$, brings it even closer to the bound.
What is remarkable -- and highly relevant to our work here -- is that the lowest vector
meson $\rho$ alone brings the energy nearly to the bound.
This is highly reminiscent of the vector dominance of the nucleon isovector electromagnetic
form factor by the $\rho$ (with a small correction from the next vector meson
$\rho^\prime$)~\cite{HRYY07,HRYY07a,HSS08,HRYY07b}.
Since the HLS Lagrangian used in our work is constructed from the five-dimensional Sakai-Sugimoto model
in a similar way, the difference being the background curvature, the inclusion of the $\rho$, and
the higher order terms in the Lagrangian makes the single skyrmion mass almost saturate the
Bogomolnyi bound.
Consequently, there is very little interaction energy left in the skyrmion made of the pion
and the $\rho$ to disturb the energy versus density curve.

We should stress that the approach to the Bogomolnyi bound by the $\rho$ -- and higher
isovector vector mesons is obstructed by the $U(1)$ vector meson $\omega$ (and its tower).
It is this $\omega$ that plays a crucial role in nuclear dynamics as will be explained below.

As emphasized and proved in Refs.~\cite{MOYHLPR12,MYOH12}, our calculation of skyrmion
properties is independent of the HLS parameter $a$.%
\footnote{In standard HLS, $a$ is a free parameter which is normally taken to be
$1 \le a \le 2$~\cite{BKUY85,BKY88,HY03a}. In free space $a \simeq 2$ is preferred, but in
a hadronic medium at high temperature and/or density, one gets $a \simeq 1$~\cite{HY03a}.
That the parameter $a$ does not figure in physical observables in this theory raises the question
as to how the vector dominance, a powerful notion in hadron physics, manifests itself in the
formalism, in particular in medium.}
The same argument can be made here, i.e., $E/B$ and the expectation value
$\langle \sigma \rangle$ are independent of the choice on the parameter $a$.
Again, as in Refs.~\cite{MOYHLPR12,MYOH12}, we could confirm the $a$ independence
of  $E/B$ and $\langle \sigma \rangle$ in the present work.

\begin{figure}[t]
\centering \includegraphics[scale=0.4]{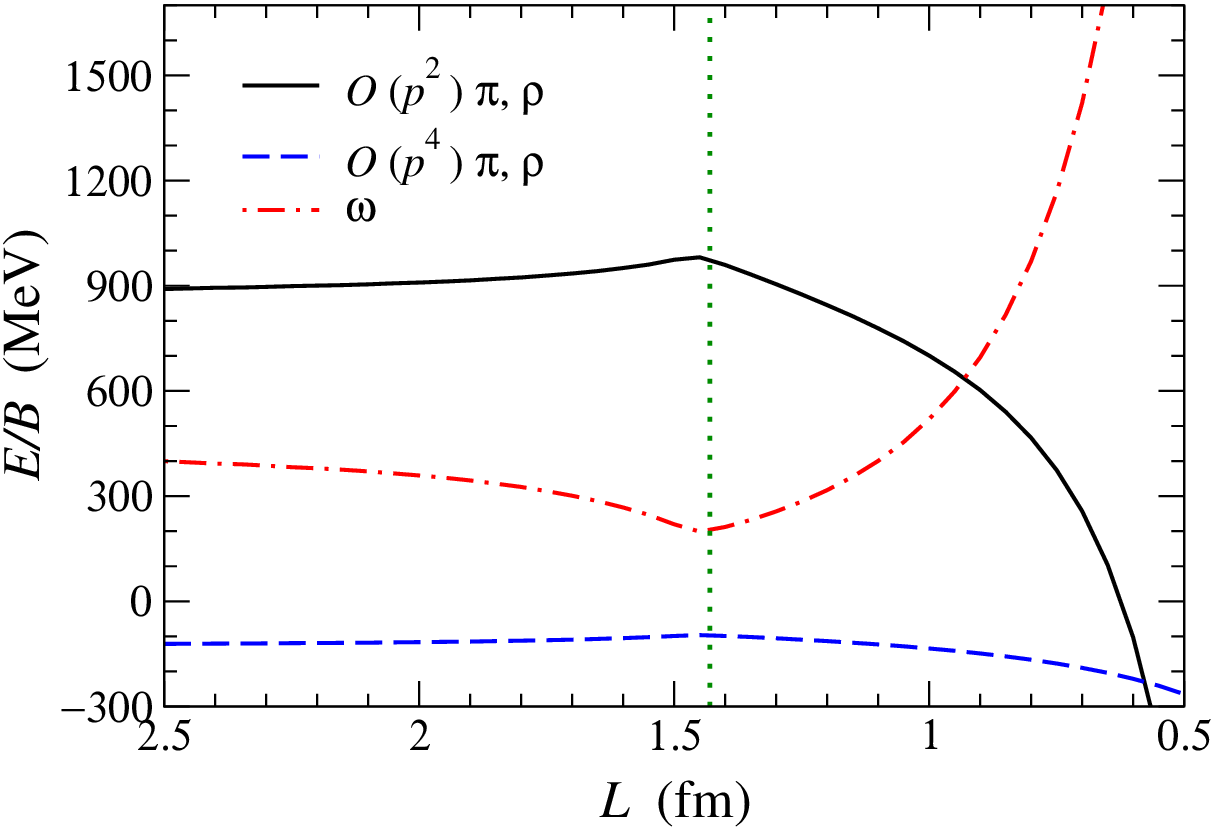}
\caption[]
{Breakdown of $E/B$. The solid, dashed, and dash-dotted lines represent
$(E/B)_{O(p^2)\pi, \rho}$, $(E/B)_{O(p^4)\pi, \rho}$, and
$(E/B)_{\omega}$, respectively. }
\label{fig:compareterm}
\end{figure}

Figure~\ref{fig:compareterm} presents the contribution of each term in $E/B$ given 
in Eq.~(\ref{EperB}).
Here, $(E/B)_{O(p^2)\pi,\rho}$, $(E/B)_{O(p^4) \pi, \rho}$, and
$(E/B)_{\omega}$ are given by the solid, dashed, and dash-dotted lines, respectively.
First of all, one can find that $(E/B)_{O(p^4)\pi,\rho}$ is negative for all density range
involved, showing clearly that the account of the entire $O(p^4)$ terms and vector
mesons brings in a sizable attraction, a feature that has not been exposed before
in the literature.
What is more striking is its weak density dependence compared to the other two
contributions.
Next, $(E/B)_{O(p^2)\pi,\rho}$ and $(E/B)_{\omega}$ show distinctly different density
dependence in the skyrmion phase and in the half-skyrmion phase.
In the single skyrmion phase, $(E/B)_{O(p^2)\pi,\rho}$ dominates over $(E/B)_{\omega}$,
with their ratio $(E/B)_{O(p^2)\pi, \rho}/(E/B)_{\omega}$ even increasing from $\sim 2$ to
$\sim 3$ as the density increases whereas in the half-skyrmion phase,
$(E/B)_{O(p^2)\pi, \rho}$ starts to decrease while $(E/B)_{\omega}$ starts to increase.
Then, at some density, the repulsive interaction due to the $\omega$ becomes dominant.

\section{Hadron properties in the FCC crystal background}
\label{sec:hadron}

\subsection{In-medium modification of meson properties}

As stressed, the power of the skyrmion-matter model is that it describes baryons and mesons
in a unified way.
This, therefore, makes it a natural framework to investigate the in-medium modification of
meson properties~\cite{LPMRV03,LPRV03a}.
The Lagrangian (\ref{eq:Lag_HLS}) itself could describe the meson dynamics in baryon free space.
If we expand the Lagrangian in terms of the fluctuating meson fields about their vacuum
values, that is, $U=1$, $\rho^a_\mu = \omega_\mu=0$, the Lagrangian (\ref{eq:Lag_HLS})
becomes
\begin{eqnarray}
\mathcal{L} &=&
\frac12 \partial_\mu \pi^a \partial^\mu \pi^a
\nonumber\\ && \mbox{}
- \frac14 \rho_{\mu\nu}^a \rho^{a\mu\nu}
+ \frac12 m_\rho^2 \rho_\mu^a \rho^{a\mu}
\nonumber\\ && \mbox{}
- \frac{1}{4} \omega_{\mu\nu} \omega^{\mu\nu}
+ \frac12 m_\omega^2 \omega_\mu \omega^\mu
\nonumber\\ && \mbox{}
+ \mbox{ interaction terms} ,
\end{eqnarray}
where we have used  $a g^2 f_\pi^2 = m_\rho^2 = m_\omega^2$.

On the other hand, what we have obtained in the previous section is the classical solutions
for $U$, $\rho_\mu^a$, and $\omega_\mu$, which are the minimum energy configurations
for a given baryon number density.
If we incorporate the fluctuations with respect to these classical solutions, then the fluctuating
fields describe the corresponding mesons and their dynamics in dense baryonic matter, which
is the main idea of Refs.~\cite{LPMRV03,LPRV03a}.
We denote the minimum energy solutions as $U_{(0)} = \xi_{(0)L}^\dagger \xi_{(0)R}^{}$,
$\rho^{a(0)}_{\mu}$, and $\omega_{\mu}^{(0)}$,
and introduce the fluctuating fields on top of the classical solutions as
\begin{eqnarray}
\xi_{L,R}^{} & = & \xi_{(0)L,R}^{} \tilde{\xi}_{L,R}^{} , \nonumber\\
\rho_\mu^a & = & \rho^{a(0)}_\mu + \tilde{\rho}^a_\mu\, , \nonumber\\
\omega_\mu & = & \omega^{(0)}_\mu + \tilde{\omega}_\mu\, , \label{eq:fieldfluct}
\end{eqnarray}
where $\tilde{\xi}^\dagger_L = \tilde{\xi}_R^{} = \tilde{\xi} = \exp (i \tau_a \tilde{\pi}_a/2f_\pi)$,
$\tilde{\rho}^a_\mu$, and $\tilde{\omega}_\mu$ stand for the corresponding fluctuating fields.
By substituting these fields into the HLS Lagrangian
and taking the space average for the background fields configurations
as denoted by $\langle \cdot \rangle$, we are led to
\begin{eqnarray}
\mathcal{L} & = &
\frac12 \left[ 1 - (a-1) \frac23 \left( 1- \left\langle \sigma^2_{(0)} \right\rangle \right) \right]
\partial_\mu \pi^a \partial^\mu \pi^a
\nonumber \\ && \mbox{}
 - \frac14 \rho_{\mu\nu}^a \rho^{a\mu\nu} + \frac{1}{2} m_\rho^2 \rho^a_\mu \rho^{a \mu}
\nonumber\\ && \mbox{}
- \frac14 \omega_{\mu\nu} \omega^{\mu\nu}
+ \frac{1}{2} m_\omega^2 \omega_\mu \omega^{\mu} + \cdots ,
\label{eq:mesoninmedium}
\end{eqnarray}
where we have considered the $O(p^2)$ terms only.
The extra factor in front of the pion kinetic term can be absorbed into the redefinition of
the pion fields as
\begin{eqnarray}
\tilde{\xi} = \exp \left( i \tau_a \tilde{\pi}^*_a/2f^*_\pi \right),
\end{eqnarray}
with
\begin{eqnarray}
\frac{\tilde{\pi}^*_a}{\tilde{\pi}_a}
= \frac{f_\pi^\ast}{f_\pi}
=  \sqrt{1 - (a-1)\frac{2}{3} \left( 1- \left\langle \sigma^2_{(0)} \right\rangle \right) } .
\label{eq:fpistar}
\end{eqnarray}
Since we have considered only the $O(p^2)$ terms, there does not appear any modification
factor for the $\rho$ meson and $f^*_\pi/f_\pi$ becomes dependent on the parameter $a$.
If we include $O(p^4)$ terms and one-loop corrections from the $O(p^2)$ terms correctly,
we believe that such an $a$-dependence will disappear or be weakened.

\begin{figure}[t]
\centering
\includegraphics[scale=0.4]{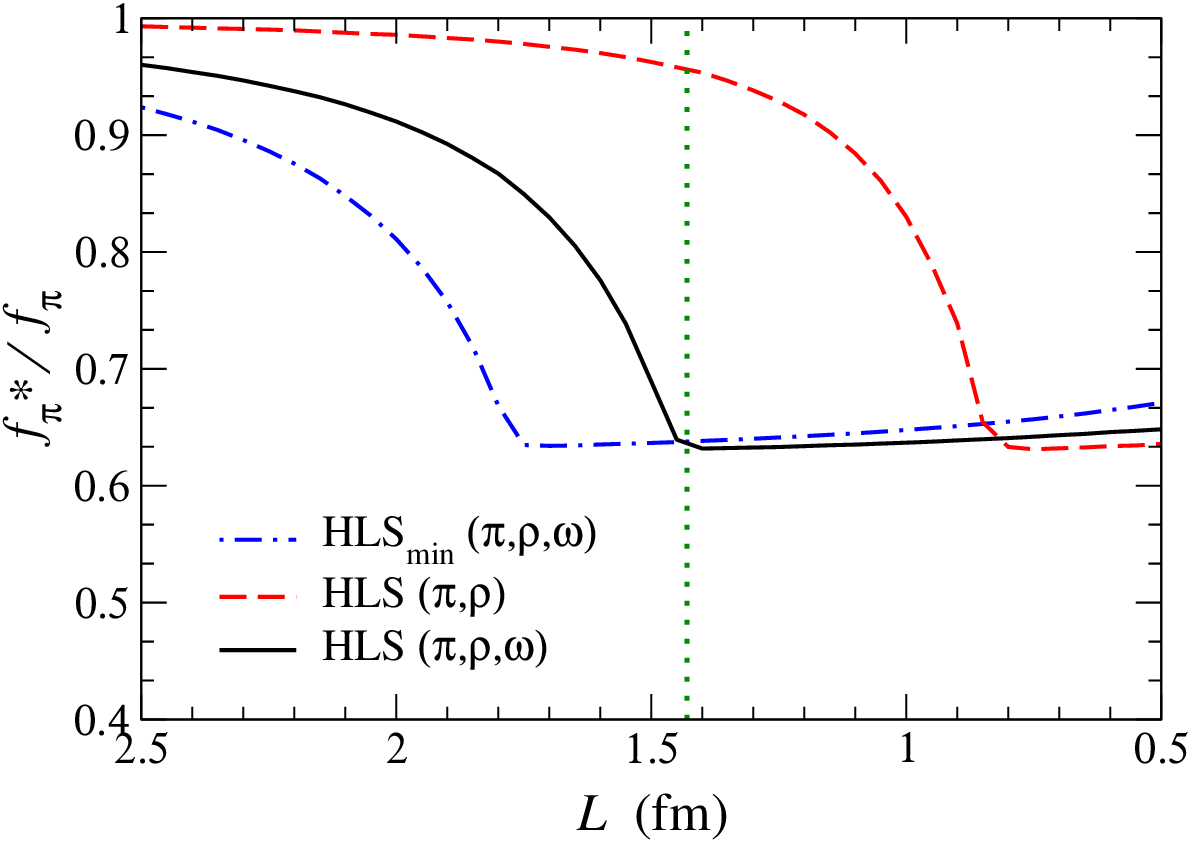}
\caption[]
{Estimates of $f_\pi^{\ast}/f_\pi$ as a function of crystal size with the skyrmion
matter background.
Notations are the same as in Fig.~\ref{fig:compareSSmin}.}
\label{fig:comparefpi}
\end{figure}

In Fig.~\ref{fig:comparefpi}, we  present $f_\pi^{\ast}/f_\pi$ as a function of the FCC box size,
where we have fixed the parameter $a$ to 2.%
\footnote{We include only the $O(p^2)$ terms, which can reasonably reproduce the meson
properties only when we take $a = 2$.}
Again, the density dependence of $f_\pi^{\ast}/f_\pi$ shows the different behavior in the
single skyrmion phase and in the half-skyrmion phase.
In the single skyrmion phase, $f_\pi^{\ast}/f_\pi$ decreases as the baryon number density
increases.
On the other hand, in the half-skyrmion phase, it stays in a nonvanishing value around 0.65.
Due to this nonvanishing $f_\pi^*$, the half-skyrmion phase was interpreted as a kind of
pseudogap phase in Ref.~\cite{LPRV03b}.

\subsection{In-medium modification of baryon properties}

In our HLS Lagrangian, all the parameters are determined by the master formula
(\ref{eq:lecshls}) given in the matter-free space.
As the simplest approximation in applying it to dense matter, we shall assume that 
the in-medium modification of all the parameters involved, such as the $O(p^4)$ terms, 
can be calculated in terms of the in-medium constants $f_\pi^{\ast}$ and $m_\rho^\ast$. 
We denote them with asterisk.
Now given these starred parameters, we can solve for the single skyrmion solution
as in Ref.~\cite{MYOH12} to obtain the in-medium modification of the
baryon mass%
\footnote{In the previous section, we have studied the skyrmion energy per baryon $E/B$.
This does not represent the quasiparticle (baryon) mass since it includes the interactions
among the baryons.}
in dense medium. To get the leading corrections, it suffices to use the $O(p^2)$ Lagrangian with $a=2$ as 
suggested by phenomenology to calculate the two in-medium constants, in which case 
the most important effect is lodged in the pion-decay constant. The effects of the $O(p^4)$ terms so calculated can then be straightforwardly incorporated with the starred quantities. In principle, we should rewrite the $O(p^4)$ terms in medium to obtain a relation which may be different from the free-space master formula, but this will be complicated considering that Lorentz invariance is broken in medium. 
Intuitively, however, we do not expect the result to be significantly modified.

\begin{figure}[t]
\centering
\includegraphics[scale=0.4]{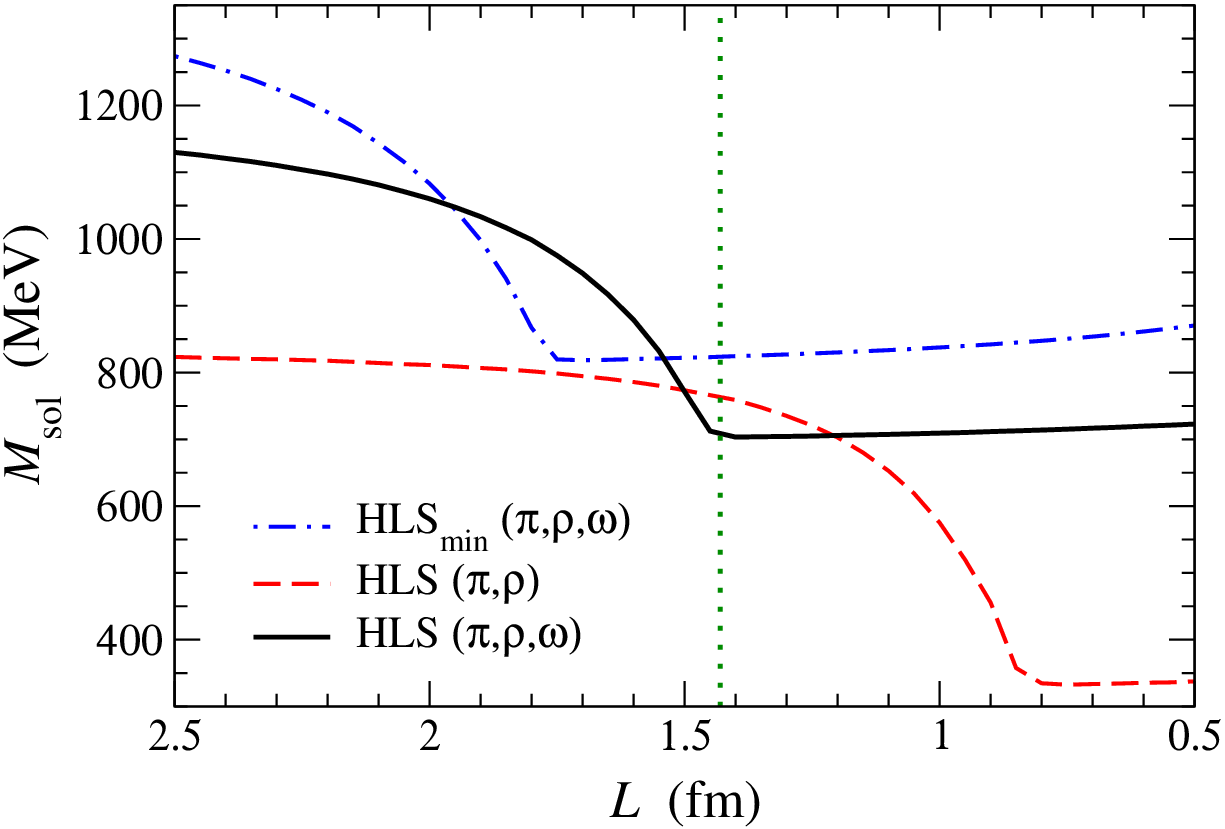}
\caption[]
{Crystal size dependence of soliton mass with the FCC crystal background.
Notations are the same as in Fig.~\ref{fig:compareSSmin}.}
\label{fig:CompareSoliMassfcc}
\end{figure}

\begin{figure}[t]
\centering
\includegraphics[scale=0.4]{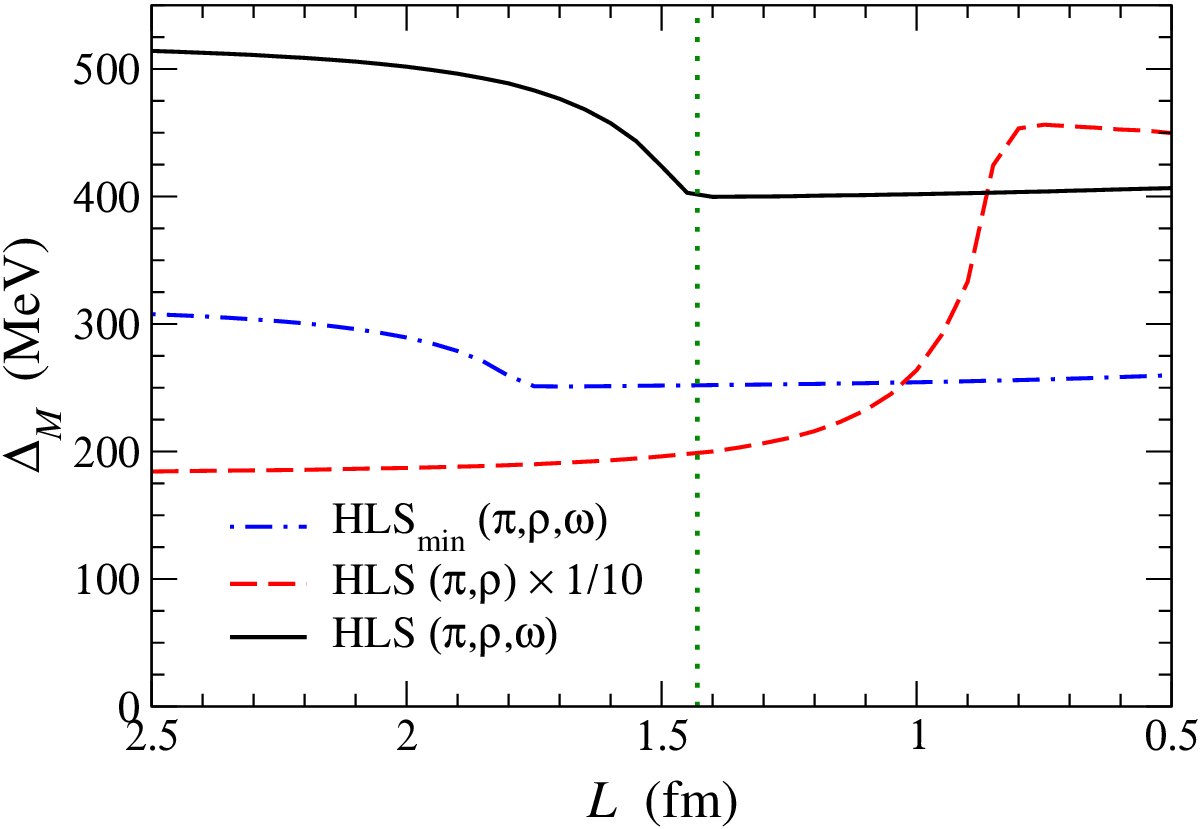}
\caption[]
{Crystal size dependence of $\Delta$-$N$ mass difference with the FCC crystal
background.
Notations are the same as in Fig.~\ref{fig:compareSSmin}.}
\label{fig:CompareSoliMassdiffcc}
\end{figure}

\begin{figure}[t]
\centering
\includegraphics[scale=0.5]{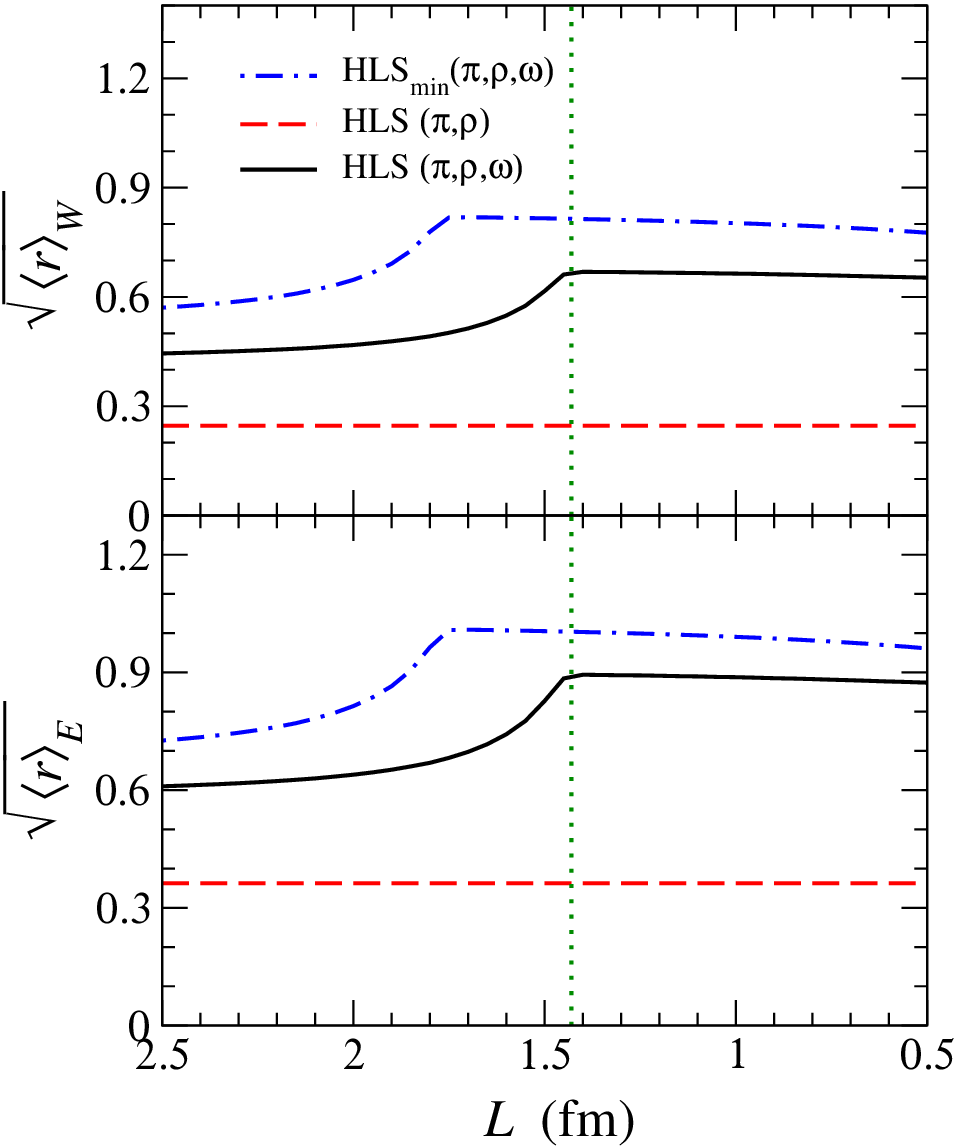}
\caption[]
{Crystal size dependence of soliton radii with the FCC crystal background.
The upper panel is for the winding number rms radius and the lower panel
is for the energy rms radius.
Notations are the same as in Fig.~\ref{fig:compareSSmin}.}
\label{fig:CompareSoliRfcc}
\end{figure}

Shown in Fig.~\ref{fig:CompareSoliMassfcc} is the in-medium single skyrmion masses 
calculated in the approximation described above as a function of the FCC size parameter $L$. 
We also present the $\Delta$-$N$ mass difference in Fig.~\ref{fig:CompareSoliMassdiffcc}
and the soliton root-mean-square (rms) radii in Fig.~\ref{fig:CompareSoliRfcc}.

Note that the behavior of the crystal size dependence of the soliton mass shown in
Fig.~\ref{fig:CompareSoliMassfcc} is the same as the behavior of $f_\pi^{\ast}$ shown
in Fig.~\ref{fig:comparefpi}.
This can be understood by that in the skyrmion model the soliton mass scales roughly 
as $M_{\rm sol} \sim f_\pi^{\ast} $, which follows simply from the scaling of the nucleon 
mass in the large $N_c$ limit as in matter-free space.
Figure~\ref{fig:CompareSoliMassdiffcc} shows that the crystal size dependence of the
$\Delta$-$N$ difference $\Delta_M$ changes a lot depending on the existence of the
$\omega$ meson.
In the calculation with the $\omega$ meson, $\Delta_M$ decreases up to $n_{1/2}^{}$ beyond
which $\Delta_M$ becomes nearly constant.
However, in the case of HLS$(\pi,\rho)$, $\Delta_M$ is unscaling up to $n_{1/2}^{}$
but, after $n_{1/2}^{}$, $\Delta_M$ increases dramatically.
Our finding that the nucleon mass, which decreases with increasing density in the skyrmion
phase, stops dropping at $n_{1/2}^{}$ and stays constant in the half-skyrmion phase
is the most significant result of this paper and, as far as we are aware, has not been obtained
in other models.
In the case of HLS$(\pi,\rho,\omega)$, the in-medium soliton mass approaches $\sim 0.6 \, M_{\rm sol}$.
As will be discussed in the next section, this finding will get some support from other
approaches, both theoretical and phenomenological, and implies consequences on the upcoming
experiments.

Similarly, the skyrmion radius is strongly influenced by the presence of the $\omega$ meson
as seen in Fig.~\ref{fig:CompareSoliRfcc}.
In the case of HLS$(\pi,\rho)$, the skyrmion radius is small and constant at varying density.
This is simply because in the absence of the $\omega$, the skyrmion is almost pointlike and
remains stable with the crystal size.
This feature was already obvious with the instantons in the hQCD model.

It is noteworthy that, while the impacts of the skyrmion--half-skyrmion transition can be
seen in all three models for the masses and splittings indicating the effect of topology
change in medium, it does not affect the skyrmion size when the $\omega$
is absent, which is in a stark contrast to the case when the $\omega$ is present.

\section{Origin of nucleon mass}
\label{matching}

The principal finding in this paper, as mentioned before, is that simulated on the crystal
lattice, the nucleon mass decreases smoothly as density increases up to $n_{1/2}^{}$
at which skyrmions fractionize into half-skyrmions and, in the half-skyrmion phase,
the dropping of the nucleon mass stops and the mass remains constant going toward
to the chiral restoration.
In the skyrmion phase with $n<n_{1/2}^{}$, chiral symmetry is broken with the order
parameter $\langle \bar{q}q \rangle \neq 0$.
Hadrons are massive apart from the Nambu-Goldstone bosons, the pions.
In the half-skyrmion phase, $\langle \bar{q}q \rangle=0$ on the average, but chiral
symmetry is not restored since hadrons are still massive and there exist pions.
There seems to be no obvious order parameter characterizing this state apart from
the presence of the half-skyrmion structure, the changeover involving a sort of topology
change.

Apart from the changes in the baryon properties discussed above, there are several other
remarkable happenings at $n_{1/2}^{}$.
For instance, in Ref.~\cite{PKR09} was found that an antikaon $K^-$ propagating in the
skyrmion matter undergoes propitious drop in its effective mass at $n_{1/2}^{}$, which
may have an influence on kaon condensation in compact-star matter.
Another phenomenon concerns the symmetry energy of nuclear matter.
At $n_{1/2}^{}$, the symmetry energy is found to develop a cusp structure which leads to
drastic modification of the nuclear tensor forces~\cite{LPR11} that contribute importantly
to the structure of asymmetric nuclei and large mass neutron stars~\cite{DKLMR12}.
We believe that all these phenomena are closely linked together and indicate the possible
source of the proton mass.

There are several indications that the nucleon mass does stay constant going toward the
chiral transition point.
It points to a possible origin of the proton mass.
The first indication comes from a phenomenological application of the skyrmion--half-skyrmion
transition to the EoS of nuclear and neutron matter~\cite{DKLMR12}.
The idea is to translate what is observed in the skyrmion description to the density dependence
of the parameters of the HLS Lagrangian, namely the masses and coupling constants scaling
with density, with which a nuclear effective field theory is constructed.
The topology change at $n_{1/2}^{}$ impacts the changes in the parameters of the EFT
Lagrangian, and the drastic effects mentioned above reflect on nuclear forces, in particular
the tensor forces.
It is observed in the EoS at high density that the nucleon mass should stabilize at $n_{1/2}^{}$ to
$m_N^* \approx 0.8 \, m_N^{}$, while the effective $\omega N N$ coupling $g^*_{\omega NN}$
can drop but very slowly in density, constrained by the constant nucleon mass.
Nature seems to be exposing an intricate interplay between $m_N^*$ and the coupling
$g^*_{\omega NN}$ or more generally the $\omega$ degree of freedom.

Another indication comes from an RG analysis of nucleon properties in HLS with baryons, i.e.,
BHLS, as scale is changed.
It is found in Ref.~\cite{PLRS13} that as one approaches the dilaton limit fixed point at which the
nonlinear sigma-model phrased in HLS transforms to a linear sigma model of
Gell-Mann-L\'evy type, the nucleon mass that falls in dilute environment stops falling at
$n_A^{}$ and then stays constant afterwards, strongly correlated with the $\omega N N$
coupling.
It is suggestive that $n_A^{}$ corresponds to $n_{1/2}^{}$ in the crystal description.

That the nucleon mass remains nonvanishing as one approaches the chiral transition point
can be captured by the nucleon parity-doublet model~\cite{DK89} with the two-component
nucleon mass (\ref{parity-doublet}) with a substantial $m_0^{} \sim (0.6-0.8)\, m_N$ that
remains nonvanishing.
In Ref.~\cite{PLRS13}, this model was used to show that the RG flow to the infrared fixed
point, i.e. dilaton-limit fixed point, does indeed indicate that there is a changeover at
some density for the nucleon mass in a close interplay with the $\omega$-nucleon
coupling.
The nonzero $m_0^{}$ can be interpreted as reflecting a symmetry intrinsic in QCD as
suggested by Glozman et al.~\cite{Gloz12,GLS12}.
This interpretation suggests a mechanism for a bulk of the mass generated differently
from the spontaneous breaking of chiral symmetry, presumably tied to chiral invariant
gluon interactions.
The interesting question is whether one cannot identify the agent for the nonzero $m_0^{}$
with one that triggers the spontaneous breaking of scale symmetry that makes the dilaton condense.

On the contrary, one can consider $m_0^{}$ representing a symmetry that is ``emerging"
from strong correlations in nuclear interactions, which has nothing to do with a fundamental
symmetry of QCD.
This is what is coming out of our skyrmion crystal picture.
In either case, it is an interesting issue in nuclear physics, intimately linked to the origin of
the nucleon mass.

\section{Remarks and discussions}
\label{sec:conclusion}

In this paper, to investigate dense baryonic matter, we simulated on crystal lattice
the HLS Lagrangian written to $O(p^4)$ in the derivative expansion with all the
parameters of the Lagrangian fixed except for the pion decay constant $f_\pi$ and
the $\rho$ meson mass $m_\rho$.
This comes about in integrating out all vector mesons in the infinite tower of vector
and axial vector mesons from a generic five-dimensional hQCD action  in a curved space plus
the Chern-Simons term, leaving for relevant degrees of freedom only the lowest vector
mesons $\rho$ and $\omega$ in a hidden local symmetric form, obtaining a master
formula that relates all $O(p^4)$ low energy constants to two constants that figure in
hQCD Lagrangian, i.e., $f_\pi$ and $m_\rho$.
The dependence on the famous $a$ parameter which figures importantly in the usual
HLS theory gauge equivalent to nonlinear sigma model, such as the KSRF-type relation
$m_V^2 = a g^2 f_\pi^2$, can be eliminated in a way analogous to a gauge parameter.
Physical quantities are independent of $a$.

The most notable findings in this work are (1) the prominent role of the $\omega$ meson
encapsulated in the homogeneous Wess-Zumino term (equivalently the $U(1)$ gauge field in
the Chern-Simons topological term in hQCD) in the structure of both elementary nucleon
and dense baryonic matter and (2) the presence of topological change from skyrmions to
half-skyrmions at a density relevant slightly above normal nuclear matter density and
higher density.

The first which plays a crucial role in giving correct structure to the nucleon controls
the location of the topology change $n_{1/2}^{}$ and accounts for the large constant
mass to the nucleon that remains nonzero at the chiral restoration which is locked to the
pion decay constant effective in the half-skyrmion phase.
This feature combined with in-medium properties of mesons introduces a dramatic change
in the nuclear tensor forces at a density above nuclear matter and gives rise to a stiff EoS
that could accommodate large neutron star masses such as the 1.97$M_\odot$ object found
recently~\cite{DKLMR12}.
Highly relevant to the forthcoming RIB machines is that the drastic
change in the nuclear tensor forces caused by the topology change we found at a density
near that of nuclear matter can have a strong consequence on the shell evolution of
asymmetric nuclei.
It has been shown that the monopole matrix element of the tensor forces follows closely
the shell evolution of single-particle states~\cite{OSHU09}.
Thus any changes in the tensor forces could be probed by looking at the shell evolution.

We should point out a few caveats and approximations made in the paper that require
further work.
The most glaring defect of this work is that the minimum of the energy density of the
system occurs at too high a density and with too large a binding energy.
One reason may be that our approximations are anchored on large $N_c$ considerations
and $1/N_c$ corrections cannot be ignored in medium.
This is a general problem with skyrmion physics which relies on the Lagrangian valid
for large $N_c$.

Perhaps more important is the mechanism that leads to a correct equilibrium density
and binding in EFT approaches to nuclear matter such as RG $V_{lowK}$~\cite{DKH11}.
There one finds that without irreducible repulsive three-body forces, the system
saturates at much higher density and with much larger binding energy, very similar to
what is found in our theory.
It is possible that the skyrmion crystal approach, while containing reducible
many-body effects, misses a mechanism that corresponds to this irreducible
three-body force effect.

Now one can ask whether the predictions made in a theory that fails to reproduce the
correct properties of normal nuclear matter can be trusted when applied to higher density.
There is no clear answer to this question for the moment.
However it may be that the quantity that depends on topology is robust.
It is seen in the calculation that although the location of $n_{1/2}^{}$ depends on
dynamical details, the existence of the skyrmion--half-skyrmion transition itself is generic.
The RG analysis of Ref.~\cite{PLRS13} provides an independent support to this result.
The issue then is where the transition takes place, and this can be ultimately checked
by experiments. There is an indication that $n_{1/2}^{}$ is near $n_0^{}$ and the phenomenology presented
in Ref.~\cite{DKLMR12} hints that $n_{1/2}^{} \sim (1.5-2.0)\, n_0$ is compatible with nature,
a region that can be accessed by the future RIB machines and is also highly relevant to
compact stars.

As already mentioned, one important missing ingredient is scalar degree(s) of freedom
in the Lagrangian.
In the Skyrme model (with pions only), the Casimir contribution to the nucleon mass
which comes at $O(N_c^0)$ is attractive and significant, of order of $\sim -0.5$~GeV.
This cannot be calculated accurately, so is generally ignored in the community.
In fact, it is the same $O(N_c^0)$ order effect that gives rise to the binding of $K^-$
to the skyrmion that describes the hyperons in the Callan-Klebanov model~\cite{CK85}.
It is very likely that this (or part of this) effect can be captured by a low-mass scalar
as in Walecka's mean field model for nuclear matter.
We have seen in our model HLS$(\pi,\rho,\omega)$ that the $\omega$-$\rho$-$\pi$ coupling
generates -- via hWZ terms -- a significant attraction in the nucleon, but there could be
further attraction that gets enhanced in dense matter.
How to implement scalars into HLS$(\pi,\rho,\omega)$ remains a problem to be worked out.
A scalar in the form of dilaton associated with spontaneously broken conformal symmetry
as exploited in Ref.~\cite{PLRS13} could provide an approach to the problem.

Among the approximations we made, we note that the crystal-size (or density) dependence
of $f_\pi$ is obtained with the $O(p^2)$ terms of HLS.
A complete calculation of this quantity up to $O(p^4)$ should include not only the tree-level
$O(p^4)$ contributions but also one-loop corrections with the $O(p^2)$ terms.
But the latter are suppressed by the $N_c$ counting.
Moreover, since we only consider fluctuations from the $O(p^2)$, both the $\rho$ and
$\omega$ meson masses are the same as their vacuum values and independent of the density.
In a complete calculation including the full $O(p^4)$ contributions which contains not only
the tree-level $O(p^4)$ terms but also one-loop corrections from the $O(p^2)$ terms, these
masses will change with density, although the changes will be suppressed by $1/N_c$.

Similarly for the density dependence of the soliton properties, only the in-medium modified
$f_\pi^{\ast}$ is considered since the other input parameter $m_\rho$ is the same as its vacuum
value in the present approximation.
Therefore, the scaling behaviors of the soliton properties agree with the large $N_c$ argument.
In a complete calculation up to $O(p^4)$ including the one-loop corrections from the
$O(p^2)$ terms, these results might be changed but again they will be suppressed by the
chiral counting or the $N_c$ counting.

\acknowledgments

A part of this work was done through the APCTP-WCU Focus Program and APCTP Topical
Research Program. We are grateful to APCTP for organizing these programs.
The work of Y.-L.M. and M.H. was supported in part by Grant-in-Aid for Scientific Research
on Innovative Areas (No. 2104) ``Quest on New Hadrons with Variety of Flavors'' from MEXT.
Y.-L.M. was supported in part by the National Science Foundation of China (NSFC) under
Grant No.~10905060.
The work of M.H. was supported in part by the Grantin-Aid for Nagoya University Global
COE Program ``Quest for Fundamental Principles in the Universe: From Particles to the Solar
System and the Cosmos'' from MEXT, the JSPS Grant-in-Aid for Scientific Research
(S) No. 22224003 and (c) No. 24540266.
The work of H.K.L. and M.R. was partially supported by the WCU project of Korean Ministry of
Education, Science and Technology (R33-2008-000-10087-0).
Y.O. was supported in part by the Basic Science Research Program through the National
Research Foundation of Korea (NRF) funded by the Ministry of Education, Science and
Technology (Grant No. 2010-0009381 and No. 2011-220-C00011).
B.-Y. P. was supported by the research fund of Chungnam National University.



\begin{thebibliography}{10}

\bibitem{MOYHLPR12}
Y.-L. Ma, Y.~Oh, G.-S. Yang, M.~Harada, H.~K. Lee, B.-Y. Park, and M.~Rho,
\newblock Phys. Rev. D \textbf{86}, 074025 (2012).

\bibitem{MYOH12}
Y.-L. Ma, G.-S. Yang, Y.~Oh, and M.~Harada,
\newblock Phys. Rev. D \textbf{87}, 034023 (2013).

\bibitem{LPMRV03}
H.-J. Lee, B.-Y. Park, D.-P. Min, M.~Rho, and V.~Vento,
\newblock Nucl. Phys. A \textbf{723}, 427 (2003).

\bibitem{PV09}
B.-Y. Park and V.~Vento,
\newblock `` Skyrmion approach to finite density and temperature,''
\newblock in \textit{The Multifaceted Skyrmions}, edited by G.~E. Brown and
  M.~Rho ( World Scientific, Singapore, 2010)

\bibitem{Skyrme62}
T.~H.~R. Skyrme,
\newblock Nucl. Phys. \textbf{31}, 556 (1962).

\bibitem{BKUY85}
M.~Bando, T.~Kugo, S.~Uehara, K.~Yamawaki, and T.~Yanagida,
\newblock Phys. Rev. Lett. \textbf{54}, 1215 (1985).

\bibitem{BKY88}
M.~Bando, T.~Kugo, and K.~Yamawaki,
\newblock Phys. Rep. \textbf{164}, 217 (1988).

\bibitem{HY03a}
M.~Harada and K.~Yamawaki,
\newblock Phys. Rep. \textbf{381}, 1 (2003).

\bibitem{SS03}
D.~T. Son and M.~A. Stephanov,
\newblock Phys. Rev. D \textbf{69}, 065020 (2004).

\bibitem{SS04a}
T.~Sakai and S.~Sugimoto,
\newblock Prog. Theor. Phys. \textbf{113}, 843 (2005).

\bibitem{SS05}
T.~Sakai and S.~Sugimoto,
\newblock Prog. Theor. Phys. \textbf{114}, 1083 (2005).

\bibitem{HRYY07}
D.~K. Hong, M.~Rho, H.-U. Yee, and P.~Yi,
\newblock Phys. Rev. D \textbf{76}, 061901 (2007).

\bibitem{HRYY07a}
D.~K. Hong, M.~Rho, H.-U. Yee, and P.~Yi,
\newblock J. High Energy Phys. \textbf{09} (2007) 063 .

\bibitem{HSS08}
K.~Hashimoto, T.~Sakai, and S.~Sugimoto,
\newblock Prog. Theor. Phys. \textbf{120}, 1093 (2008).

\bibitem{RSZ09}
M.~Rho, S.-J. Sin, and I.~Zahed,
\newblock Phys. Lett. B \textbf{689}, 23 (2010).

\bibitem{KMS12}
V.~Kaplunovsky, D.~Melnikov, and J.~Sonnenschein,
\newblock J. High Energy Phys. \textbf{11} (2012) 047.

\bibitem{HMY06}
M.~Harada, S.~Matsuzaki, and K.~Yamawaki,
\newblock Phys. Rev. D \textbf{74}, 076004 (2006).

\bibitem{HMY10}
M.~Harada, S.~Matsuzaki, and K.~Yamawaki,
\newblock Phys. Rev. D \textbf{82}, 076010 (2010).

\bibitem{Song00}
C.~Song,
\newblock Phys. Rep. \textbf{347}, 289 (2001).

\bibitem{PLRS13}
W.-G. Paeng, H.~K. Lee, M.~Rho, and C.~Sasaki,
\newblock arXiv:1303.2898.

\bibitem{DK89}
C.~E. Detar and T.~Kunihiro,
\newblock Phys. Rev. D \textbf{39}, 2805 (1989).

\bibitem{DKLMR12}
H.~Dong, T.~T.~S. Kuo, H.~K. Lee, R.~Machleidt, and M.~Rho,
Phys. Rev. C \textbf{87}, 054332 (2013).

\bibitem{Gloz12}
L.~\mbox{Ya}. Glozman,
\newblock Acta Phys. Polon. B Proc. Supp. \textbf{6}, 245 (2013).

\bibitem{GLS12}
L.~\mbox{Ya}. Glozman, C.~B. Lang, and M.~Schrock,
\newblock Phys. Rev. D \textbf{86}, 014507 (2012).


\bibitem{KS88}
M.~Kugler and S.~Shtrikman,
\newblock Phys. Lett. B \textbf{208}, 491 (1988).

\bibitem{KS89a}
M.~Kugler and S.~Shtrikman,
\newblock Phys. Rev. D \textbf{40}, 3421 (1989).

\bibitem{PRV03}
B.-Y. Park, M.~Rho, and V.~Vento,
\newblock Nucl. Phys. A \textbf{736}, 129 (2004).



\bibitem{Sutcliffe10}
P.~Sutcliffe,
\newblock J. High Energy Phys. \textbf{1008}, 019 (2010).

\bibitem{Sutcliffe11}
P.~Sutcliffe,
\newblock J. High Energy Phys. \textbf{1104}, 045 (2011).

\bibitem{HRYY07b}
D.~K. Hong, M.~Rho, H.-U. Yee, and P.~Yi,
\newblock Phys. Rev. D \textbf{77}, 014030 (2008).

\bibitem{LPRV03a}
H.-J. Lee, B.-Y. Park, M.~Rho, and V.~Vento,
\newblock Nucl. Phys. A \textbf{726}, 69 (2003).

\bibitem{LPRV03b}
H.-J. Lee, B.-Y. Park, M.~Rho, and V.~Vento,
\newblock Nucl. Phys. A \textbf{741}, 161 (2004).

\bibitem{PKR09}
B.-Y. Park, J.-I. Kim, and M.~Rho,
\newblock Phys. Rev. C \textbf{81}, 035203 (2010).

\bibitem{LPR11}
H.~K. Lee, B.-Y. Park, and M.~Rho,
\newblock Phys. Rev. C \textbf{83}, 025206 (2011),
\newblock \textbf{84} \textit{ibid} \textbf{84}, 059902(E) (2011).

\bibitem{OSHU09}
T.~Otsuka, T.~Suzuki, M.~Honma, Y.~Utsuno, N.~Tsunoda, K.~Tsukiyama, and
  M.~Hjorth-Jensen,
\newblock Phys. Rev. Lett. \textbf{104}, 012501 (2010).

\bibitem{DKH11}
H.~Dong, T.~T.~S. Kuo, and J.~W. Holt,
arXiv:1105.4169.

\bibitem{CK85}
C.~G. Callan and I.~Klebanov,
\newblock Nucl. Phys. \textbf{B 262}, 365 (1985).

\end{thebibliography}
\end {document}